# Conjugate adaptive optics with remote focusing for three-dimensional focusing through scattering media

Xiaodong Tao,[1,*] Tuwin Lam,[1] Bingzhao Zhu,[2] Qinggele Li,[1] Marc R. Reinig,[1] Joel Kubby[1]

[1]*W.M. Keck Center for Adaptive Optical Microscopy, Jack Baskin School of Engineering, University of California, Santa Cruz, CA 95064, USA*
[2]*State Key Laboratory of Modern Optical Instrumentation, College of Optical Science and Engineering and the Collaborative Innovation Center for Brain Science, Zhejiang University, Hangzhou, Zhejiang 310027, China*
*\*Corresponding author: xiaod.tao@gmail.com*

**Abstract.** The small correction volume for conventional wavefront shaping methods limits their applications in biological imaging through scattering media. We demonstrate large volume wavefront shaping through a scattering layer with a single correction by conjugate adaptive optics and remote focusing (CAORF). The remote focusing module can keep the conjugation between the AO and scattering layer during three-dimensional scanning. This new configuration provides a wider correction volume by the best utilization of the memory effect in a fast three-dimensional laser scanning microscope. Our results show that the proposed system can provide 10 times wider axial field of view compared with a conventional conjugate AO system when 16,384 segments are used on a spatial light modulator. We also demonstrated three-dimensional fluorescence imaging and multi-spot patterning through a scattering layer.

**Keywords**: Adaptive optics, Imaging through turbid media, Conjugate adaptive optics.

## 1. INTRODUCTION

High-resolution optical imaging through scattering media is extremely important for noninvasive biological imaging. Most of optical imaging systems rely on the ballistic light through the biological sample to form an image. As the imaging depth increases, multiple scattering becomes a dominant factor limiting the image depth. One way to minimize the scattering effect is to use long wavelength light since the longer wavelength gives less scattering [1, 2]. The use of longer wavelength excitation in multiphoton imaging systems has been successfully demonstrated for in-vivo imaging of live animals with extended penetration depth [1-5]. However, special laser systems with low-repetition rate, higher-pulse energy and additional dispersion compensation are required to achieve an acceptable signal-to-noise-ratio with minimal photodamage or photobleaching. In addition, the increase of water absorption can limit the imaging depth when some common laser wavelengths such as 1550 nm are used [3]. Another promising way is to shape the wavefront using adaptive optics (AO). AO has been used to correct low-order aberrations from refractive index mismatch in biological samples [6-12]. The wavefront shaping methods for compensation of randomly distributed wavefront has been investigated extensively in recent years [13]. The complex wavefront from scattering media can be recorded directly using interferometric techniques and the desired optical field is replayed by optical or digital phase conjugations [14-16]. The desired compensation wavefront can also be estimated by feed-back optimization algorithms or transmission matrix measurements [17-22]. The first demonstration of focusing through scattering media by Vellekoop et al. [17] shows its potential applications in fluorescence laser scanning microscopy. The lateral or axial scanning of the focus behind the scattering layer was further demonstrated utilizing the memory effect by several groups [23-29]. Lateral scanning is realized by either rotating the beam around the scanning layers or by adding tip/tilt to the compensation phase. Adding additional quadratic phase or applying wavefront measurement at different focal planes provides the additional axial scanning. However, the configuration using a rotating beam around sample is not practical for a conventional laser scanning system, where the beam is often translated at the object plane using a high-speed scanner. Furthermore, the slow lateral or axial scanning by updating the phase on a spatial light modulator (SLM) cannot meet the temporal requirement for in vivo biological imaging. Recently a conjugate AO (CAO) multiphoton system has been demonstrated for high-resolution in-vivo imaging of neurons through a mouse skull [30]. The scattering from the mouse skull is compensated by a segmented deformable mirror that is conjugated with the skull. Since the conjugation plane moves during axial scanning, the results show a limited axial field of view (FOV) that is only tens of microns.

In this paper, we take advantages of CAO and remote focusing (CAORF) to achieve three-dimensional (3D) scanning through a scattering layer with a single correction. First, the transmission matrix is measured iteratively at the target point when the SLM conjugates with the scattering layer. CAORF provides a large lateral FOV due to the memory effect. During axial scanning, the remote focusing (RF) module shifts the focal plane without changing the conjugation. CAORF can achieve a large correction volume without updating the phase on the SLM. It can be easily integrated into a laser scanning fluorescence microscope, such as a multiphoton microscope, for in-vivo imaging through scattering tissue. High-speed volumetric imaging can be achieved by using a fast RF module, which will benefit a wide range of applications, such as functional imaging of neurons through scattering layers in live animals. In addition, the iterative transmission matrix measurement in this system makes it easy to use a fluorescent guide-star in the sample [10, 31].

## 2. CONCEPT

Three configurations of the AO laser scanning systems are shown in Fig. 1 (a-c). The conventional pupil AO projects the phase on the SLM to the pupil plane of the objective, as shown in Fig. 1 (a). The conjugate plane shifts laterally and axially during 3D scanning. This setup is efficient to compensate low-order refractive aberration, where the FOV is limited by the transverse correlation length $\sigma_x$ of the distorted wavefront. For highly scattering media, where the transverse correlation length of the distorted wavefront is much smaller than the SLM segment size, this configuration gives very limited lateral and axial FOV. The lateral FOV is determined by [28]:

$$X_{FOV} \sim 2\sigma_{SLM} \tag{1}$$

where $\sigma_{SLM}$ is the SLM segment size on the conjugate plane. When the scattering layer or the objective shifts along the Z-axis, $\Delta_Z$, the segment phase on the scattering layer is rescaled along the optical axis because of the beam convergence, as shown in Fig. 1 (a). The effective area in each channel for scattering compensation decreases with the axial shift. The ratio of the intensity at the $\Delta_Z$ plane to that at the initial conjugate plane becomes

$$\boldsymbol{\eta}_{\Delta z} \sim \left(\frac{\sum_n \alpha_{n,\Delta z}}{N}\right)^2 \tag{2}$$

where $\boldsymbol{\alpha}_{n,\Delta z}$ is the ratio of the effective area to the segment area at the n[th] segment. At the conjugate plane, $\boldsymbol{\alpha}_{n,\Delta z} = 1$. N is the total number of the segments.

The CAO projects the SLM onto the scattering layer as shown in Fig. 1(b). It provides a wider lateral FOV, determined by the optical 'memory effect' for the speckle correlations [23, 24]. The lateral FOV obtained as the full width at 1/e² of the intensity correlation is given by $X_{FOV,e^2} \sim 2.686 Z\lambda/\pi L$, where L is the thickness of the scattering layer [32]. Z is the distance between the scattering layer and the focal plane. However, this configuration does not maintain conjugation during axial scanning when translating the sample. The axial FOV is same as the pupil AO system, determined by Eq. (2).

To extend the axial FOV, a new method, integrating an RF module into a CAO system, also called CAORF, is proposed as shown in Fig. 1 (c). The purpose of the RF module is to shift the focal plane without moving the sample or the objective lens. This maintains the conjugation between the wavefront corrector and the scattering layer during 3D scanning. Now the axial FOV becomes [25]

$$Z_{FOV} \sim 2z\, X_{FOV}/D \tag{3}$$

where D is the diameter of the beam on the scattering layer. As can be seen, the axial FOV is not related to the SLM segment number or size. The CAORF can provide a wide axial FOV even when a large number of segments are used. Eq. (3) does not consider the change of the effective area during 3D scanning. Analysis that is more rigorous is made by numerically calculating the memory-effect intensity correlations with both lateral and axial shifts.

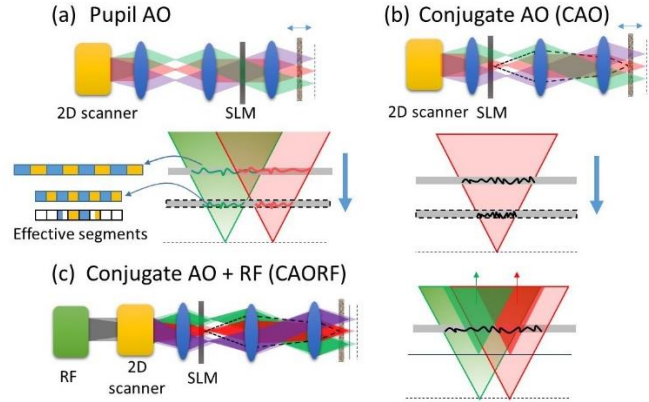

Fig 1. A schematic diagram of AO laser scanning systems. (a) Pupil AO. The SLM is conjugated with the pupil plane. The projected wavefront at the scattering layer moves with the scanning beam. The effective number of segments decreases with the axial shift. (b) Conjugate AO (CAO). The SLM is located at the plane that is conjugate to the scattering layer. The projected phase changes with the axial shift. (c) CAO and remote focusing (CAORF). The projected phase on the scattering layer does not change during three-dimensional scanning.

## 3. EXPERIMENTAL SETUP

Fig. 2 shows the system setup to evaluate this new configuration. A solid-state laser (488 nm, LuxX 488-60, Omicron) is used as the light source for both FOV evaluation and fluorescence imaging. It is collimated by an achromatic lens L1 (f=150mm). As a proof-of-concept system, an achromatic lens L2 (f=100mm) installed on a three-axis motorized stage works as the RF and two-dimensional (2D) scanning module. Lateral scanning on the sample is realized by the translation of the lens along the X-Y axis. Translation along the Z-axis shifts the focal plane without moving the objective lens or the sample. For real biological imaging applications, a fast 2D scanner and RF module can be used instead. The conjugate AO module includes a reflective SLM (RCL-2500, Holoeye) and a folding mirror M1 installed on a translation stage. Lens L3 (f=100mm) and an objective lens O1 (10x, NA0.3, LEITZ) projects the phase on the SLM to the scattering layer. The conjugation planes are aligned precisely by the translation stage without changing the final focal plane. Light from the sample is collected by another objective O2 (20x, NA 0.5, LEITZ) and then focused on a CCD camera (Genie M1400-1/2, Teledyne DALSA) by a lens L4 (f=150mm). For fluorescence imaging, an emission filter (FF01-542/27, Semrock) is put in the detection path, which is not shown in Fig. 1. Single sided tape (Scotch, 3M) attached on a coverslip is used as the scattering layer. The sample is mounted on a translation stage for precise adjustment.

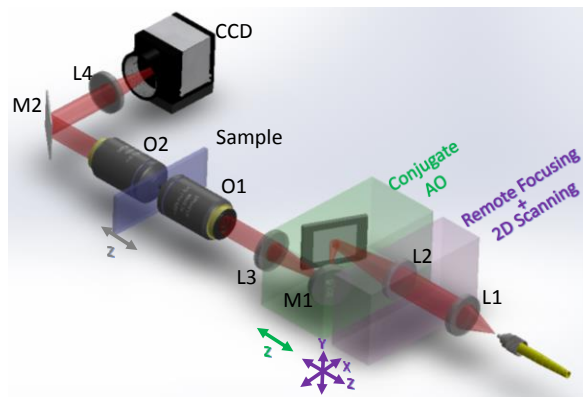

Fig. 2. Schematic of the experimental setup. A lens (L1) collimates the laser beam from a 488nm solid-state laser. A lens (L2) installed on a three-axis translation stage works as a remote focusing (RF) and two-dimensional (2D) scanning module. The conjugate AO module consists of a spatial light modulator (SLM) and a folding mirror (M1), which are installed on a single-axis translation stage. The lens (L3) and an objective (O1) image the SLM plane onto the scattering lay. The light is collected by another object (O2) and focused by a lens (L4) on a CCD camera.

## 4. ITERATIVE WAVEFRONT SHAPING

An interferometric focus is generated at the backside of the scattering layer by measuring the transmission matrix (TM) iteratively at the target position [21]. To utilize the full aperture and increase the signal-to-noise-ratio during the TM measurement, the whole aperture is divided into two groups by random selection of half of the segments, as shown in Fig. 3 (a). Two measurement steps for those two groups are carried out. In the first step, the first group is selected as the active group and each mode defined by the Hadamard basis is displayed on the SLM. The phase of each mode at the output is measured by updating the phase of the other group, the reference group, from 0 to $2\pi$ using a four-phase method [18]. After the first measurement, the optimal phase is updated on the first group as shown in Fig. 3 (b). Then the same procedure is performed on the other group to achieve the final phase mask as shown in Fig. 3 (c). Figure 3 (d-f) shows the focus before correction, and after the first and second steps, respectively, when a total of 1024 segments are used for the TM measurement. The intensity is doubled after the second correction, as expected. Compared with the direct phase measurement using optical phase conjugation [14, 15], this iterative TM measurement method is more practical for fluorescence imaging, where the emission light is incoherent. The measurement time is limited by the frame rate of the SLM. Currently the total measurement time for 1024 channels is 135 seconds. Since only a single correction is required for 3D scanning, it is tolerable when imaging through a static scattering layer. Potentially the measurement time could be reduced to less than half a second when a high speed segmented deformable mirror or a digital micro mirror device is used [33-35].

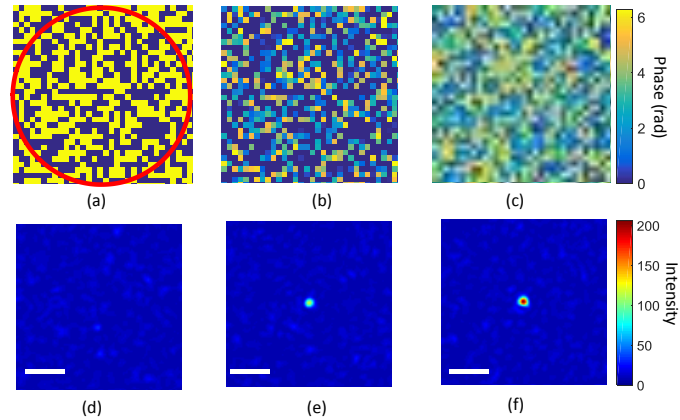

Fig. 3. Iterative wavefront shaping. (a) The whole aperture is randomly divided into two groups of segments. (b) The optimal phase of the first group after the transmission matrix measurement. (c) After the optimal phase for the second group of segments is achieved, the combined phase mask is displayed on the SLM. The image before correction, after the first correction and after the second correction are shown in (d), (e) and (f) respectively. The scale bar is 5 μm.

## 5. EXPERIMENTAL RESULTS

### A. 3D Focus through Scattering Layers

To evaluate the ability of 3D focusing through the scattering layer, a coverslip with a strip of scattering tape is placed 2 mm from the focal plane of lens O1. Then lens O2 is moved accordingly to focus on the scattering layer. The conjugate AO module is then adjusted to the conjugate plane of the scattering layer. The correct conjugation is verified by imaging both the SLM and the scattering layer on the camera. Then lens O2 moves back to the focal plane of lens O1 for monitoring the focal spot. The transmission matrix measurement is performed to generate an interferometric focus behind the scattering layer. To evaluate the focal spot in 3D space, the beam is scanned in the *X-Z* plane by the RF and 2D scanning modules, as shown in Fig. 4 (a). To monitor the focus at different depths, the objective lens O2 is refocused at each new depth. To compare with the results without RF, the scattering layer is moved along the *Z*-axis as shown in Fig. 4(b). The normalized peak intensity of the focal spot along the *X-Z* plane with RF and without RF is shown in Fig. 4 (c) and Fig. 4(d), respectively. The CAORF gives a much slower intensity drop with an axial shift in comparison to one without RF. The simulation of the CAORF shows a similar intensity drop with the axial shift. Fig. 4 (e) shows the normalized peak intensity of the spot at the center of the field at different axial shifts. The red and blue dashed lines indicate the measurement data for the configurations with and without RF, respectively. The axial FOV measured by the full-width-half-maximum (FWHM) increases from 0.1 mm to 0.43 mm when RF is applied. The simulation result is indicated by the blue line. The deviation from the experimental result could be caused by misalignment and conjugation errors from the tilted SLM.

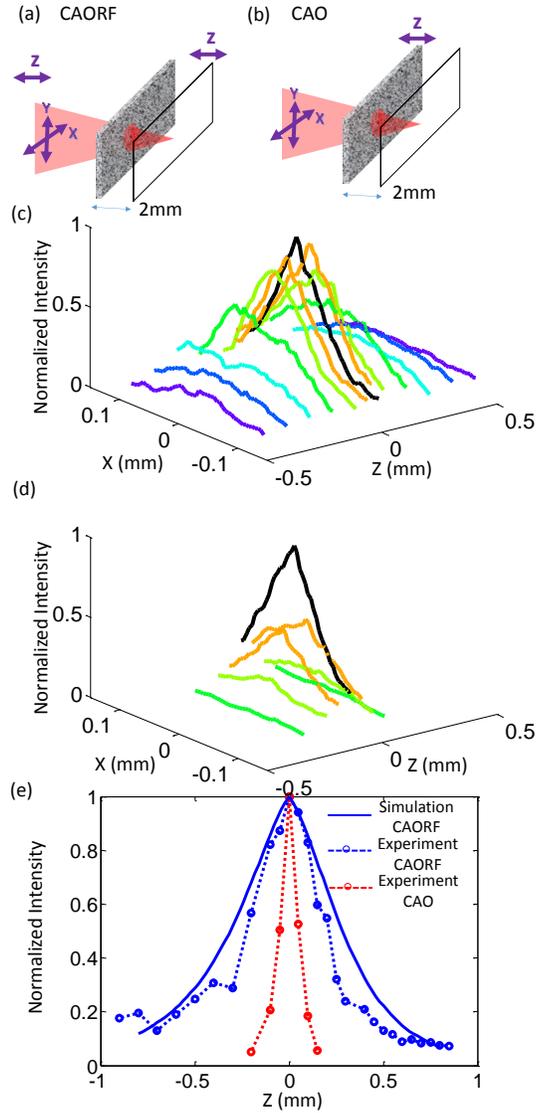

Fig. 4. Field of view analysis for 3D scanning for CAORF (a) and CAO without RF (b). The normalized intensity of the focal spot scanning along the X-Z plane for these two configurations are shown in (c) and (d), respectively. (e) The normalized intensity of the focal spot along the center of the field versus the axial shift. The red and blue dashed curves show measurement results for CAORF and CAO, respectively. The simulation of CAORF (a) is indicated by the blue curve.

Another advantage of CAORF is the insensitivity of the axial FOV to the number of segments on the SLM. A full TM is often comprised of millions of elements. Even a small fraction of the full matrix still requires thousands of channels on the SLM. For a large number of segments, the focus will be lost even with a small conjugation error caused by the focus shift. Using RF to maintain the conjugation between the SLM and the scattering layer, the axial FOV is only determined by the memory effect. Figure 5 shows the normalized intensity of the focus at various focal distances with 1024, 4096 and 16,384 segments on the SLM. The measurements with CAORF are indicated by the triangles. The results with different numbers of segments have a similar intensity drop with the axial shift, which is around 200 µm at the half-width-half-maximum (HWHM). However, without RF, the configuration with more segments gives a much smaller axial FOV. The HWHMs for 1024, 4096 and 16,384 segments drop to 19 µm, 27 µm and 40 µm, respectively. Figure 5 also shows the simulations in different situations based on Eq. 2. The difference between the measurement and simulations could be caused by the alignment error or the conjugation error from a tilted SLM. The results show the improvement of the axial FOV by 5, 7 and 10 times for 1024, 4096 and 16,384 segments, respectively.

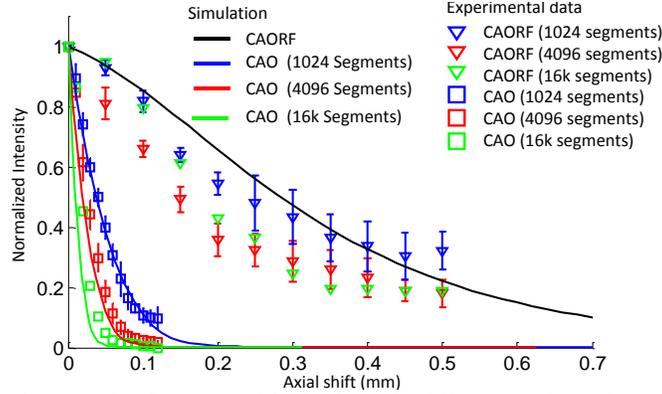

Fig. 5. Effect of the axial shift on the normalized intensity of the focal spot for different numbers of segments with CAORF and CAO. The measurements with CAORF and CAO are indicated by triangles and squares. The solid curves shows the simulation results in each case. The results for 1024, 4096 and 16,384 segments are indicated by blue, red and green, respectively.

The theoretical FWHM of the focal spot for a circular aperture defined by $0.51\lambda/NA$ is 0.84 nm with an NA of 0.3 and a wavelength of 488nm in the present system. The images of the focal spot at different focal planes are shown in Fig. 6 (a). The experimental FWHM measurements of the focal spot with the focus shifts are shown in Fig. 6 (b). The near diffraction limit focal spot (FWHM=0.94 μm) at the original focal plane (Z=0 μm) is achieved. The deviation from the theoretical calculation maybe caused by system misalignment. The spot size increases with depth because of background noise. The spot size is increased by 1.2 times at a depth of 0.5mm with CAORF. However, a similar spot size is achieved only at the depth of 0.06mm for the conventional CAO system.

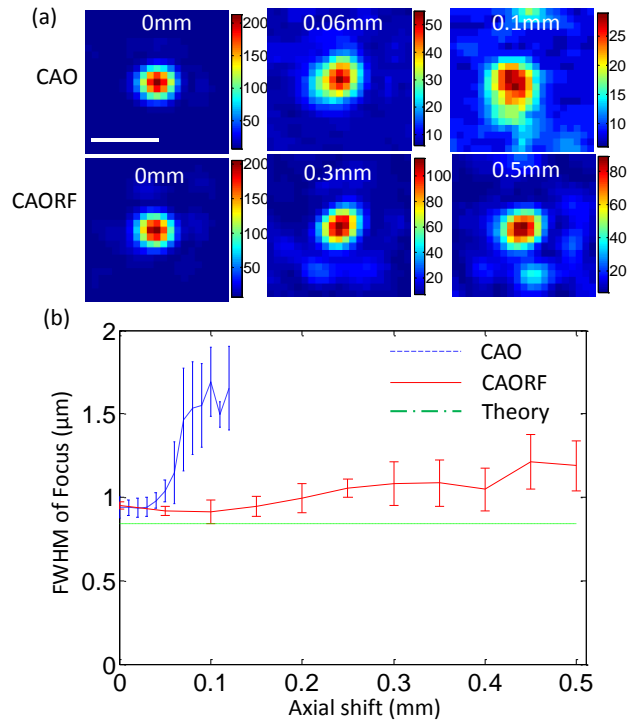

Fig. 6 (a) Images of the focal spot at different focal planes for CAO and CAORF. (b) The full-width-half-maximum (FWHM) of the focal spot at different focal distances. The measurements for CAO and CAORF are indicated by blue and red lines. The theoretical resolution is indicated by the green line. The scale bar is 2μm.

## B. 3D Multi-Spot Patterning

An advantage of using the SLM as a wavefront corrector is the ability to generate an arbitrary pattern through the scattering layer [17, 29]. To generate a 3D pattern, the phase often needs to be measured at different $Z$ planes because of the limited axial FOV. Sequential generation of the multi-spot pattern at different focal planes requires an update of the phase masks on the SLM, which limits the temporal resolution of the system. Here we show that CAORF can scan 3D patterns while scanning along the *Z-axis* with RF.

Fig. 7 (a) shows the configurations of the experiments. The whole aperture is divided into two groups of channels by random selection of the segments. The transmission matrix for each group is measured at two different focal planes separated by 120 μm. The combined phase masks generate two patterns at the two focal planes at 0 μm and 120 μm respectively. The RF module scans two patterns simultaneously from 0 to 120 μm and 120 μm to 240 μm, respectively. The 3D patterns at different focal planes are confirmed by the images from the camera as shown in Fig. 7 (b). The detector objective lens translates along the optical axis to focus at each plane.

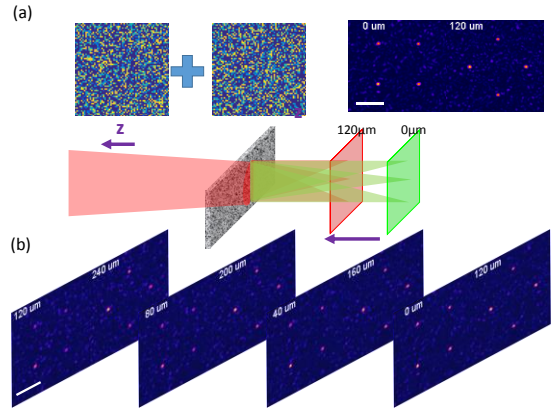

Fig. 7. Axial scanning 3D patterns with RF. (a) Two patterns at two focal planes (Z=0 μm and 120 μm) are generated simultaneously by combing two phase masks. The remote focusing module shifts the focal plane with a range of 120 μm. (b) The corresponding images at different focal planes are captured by the CCD camera. The scale bar is 10 μm.

### C. 3D Fluorescent Imaging

To test the systems performance for 3D fluorescent imaging through a scattering layer, a coverslip with 1.1 μm diameter fluorescent beads (FluoSpheres, Invitrogen) was placed at a 2mm distance behind the scattering tape. The emission light is collected by a detector after an emission filter as shown in Fig. 8 (a). In the present system, the CCD camera works as the photodetector. To generate the final 2D image, the intensity at each scanning position is calculated as the sum of a large region of interest (100×100 pixels) centered at the focal spot on the CCD camera. In total 200×200 points are collected, which produced the final image with 200×200 pixels. The image before wavefront correction is shown in Fig. 8(b). It is hard to observe the fine structure because of scattering. Fig. 8 (c) shows the images when the scattering layer moves from 0 μm to 150 μm towards the sample. This is similar to axial scanning by moving the sample. At a depth of 0 μm, the structure of the beads can be observed after correction when the SLM is conjugated with the scattering laser. The image contrast degrades quickly with an increase of the axial shift. At an axial shift of 150 μm, the fine structure cannot be observed. In the case of CAORF, the focal plane is moved toward the scattering layer by the RF module. The sample also moves the same distance using a translation stage. Fig. 8 (d) shows images at depths from 100 μm to 500 μm. As can be seen, the fine structure can be still observed even at a depth of 500 μm.

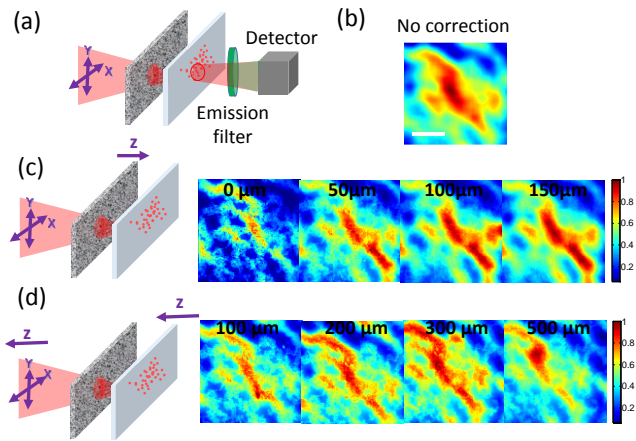

Fig. 8. Comparison of 3D fluorescence imaging with CAO and CAORF. (a) The sample is placed 2mm behind the scattering layer. The beam is translated in the *X-Y* plane by the 2D scanning module. The emission light is collected by the detector, which gives the intensity information for each scan position to generate a 2D image. (b) The image without wavefront shaping. (c) The images at depths of 0 μm,

50 µm, 100 µm and 150 µm without RF. The depth is adjusted by moving the scattering layer. (d) The images at depths of 100 µm, 200 µm, 300 µm and 500 µm with CAORF. The depth is adjusted by the remote focusing module. . The scale bar is 50µm.

## 6. DISCUSSION AND CONCLUSIONS

By the combination of conjugate AO and RF, the CAORF system maintains the conjugation between the scattering layer and the SLM during axial scanning, which can provide an extended axial FOV. No additional update of phase on the SLM is required. The axial scanning speed is only limited by the remote focusing module. As a proof-of-concept system, a single lens on a three-axis translation stage was used for the RF and 2D scanning. It can easily be replaced by a high-speed RF module, such as an acousto-optic deflector (AOM) [36, 37], an electrically tunable lens [38, 39], an ultrasound lens [40] or a reference lens with an axial scan mirror [41]. The lateral scanning speed can be improved by using a high-speed resonant scanner or an AOM [42]. In this paper, we also demonstrate three-dimensional multi-spot formation through a scattering layer. The results show its potential application in multiplane structural imaging with the ability for scattering and refractive aberration compensation [43].

Although we only show the compensation of scattering, the present system can also correct low-order refractive aberration as well. Comparing with existing CAO microscopy systems [44, 45, 46], the proposed system can achieve larger axial FOV by minimizing the conjugation error during axial scanning.

In addition to the memory effect, the axial FOV is also limited by system aberration from the objective lens, which is not discussed in this paper. Theoretically, the wavefront for the system aberrations could be calibrated and added on to the SLM during wavefront shaping for scattering compensation. For a thick sample with multiple scattering, a numerical study has shown that multi-conjugate AO, originally used in astronomy, could be a promising solution [47]. The proposed configuration in this paper could also benefit these systems to achieve a large FOV with near diffraction-limited resolution.

In conclusion, we demonstrated a large volume wavefront shaping method by combining CAO and RF. This configuration minimizes the conjugation error during axial scanning and provides a much larger axial FOV. Compared with a system without RF, the experiments show a 10x axial FOV improvement for 16,384 input channels on the SLM when a single layer of Scotch tape is used as a scattering layer. We demonstrated three-dimensional multi-spot patterning through the scattering layer. The result of three-dimensional fluorescent imaging shows its ability to extend the axial FOV for laser scanning microscopy.

**Funding.** The results presented herein were obtained at the W. M. Keck Center for Adaptive Optical Microscopy (CfAOM) at University of California Santa Cruz. The CfAOM was made possible by the generous financial support of the W. M. Keck Foundation. This material is based upon work supported by the National Science Foundation under Grant Numbers 1353461 and 1429810. Any opinions, findings, and conclusions or recommendations expressed in this material are those of the authors and do not necessarily reflect the views of the National Science Foundation.